# Mars, a Post-Habitable Planet?

**Authors:** Matteo Crismani matteo.crismani@csusb.edu, Richard Cartwright, Michael Chaffin, Sara Faggi, Stephanie Milam, Geronimo Villanueva

1. **Introduction:** Mars provides a critical analog to once habitable exoplanets that have since lost their surface liquid water. The current atmospheric state of Mars retains the chemical fingerprints of that transition, including isotopic signatures of atmospheric escape and climate evolution. As the closest accessible example of a terrestrial world with definitive evidence for once supporting liquid water on its surface, Mars presents a unique opportunity to test hypotheses about planetary habitability and atmospheric evolution in a spatially and temporally resolved way.

   Observations of Mars with HWO will inform the interpretation of exoplanet spectra by helping to constrain climate models, water loss estimates, and understanding of atmospheric escape. Such observations will also support our understanding of the variability and physical processes that drive Martian climate today, including cloud and dust storm formation, as well as atmospheric circulation.

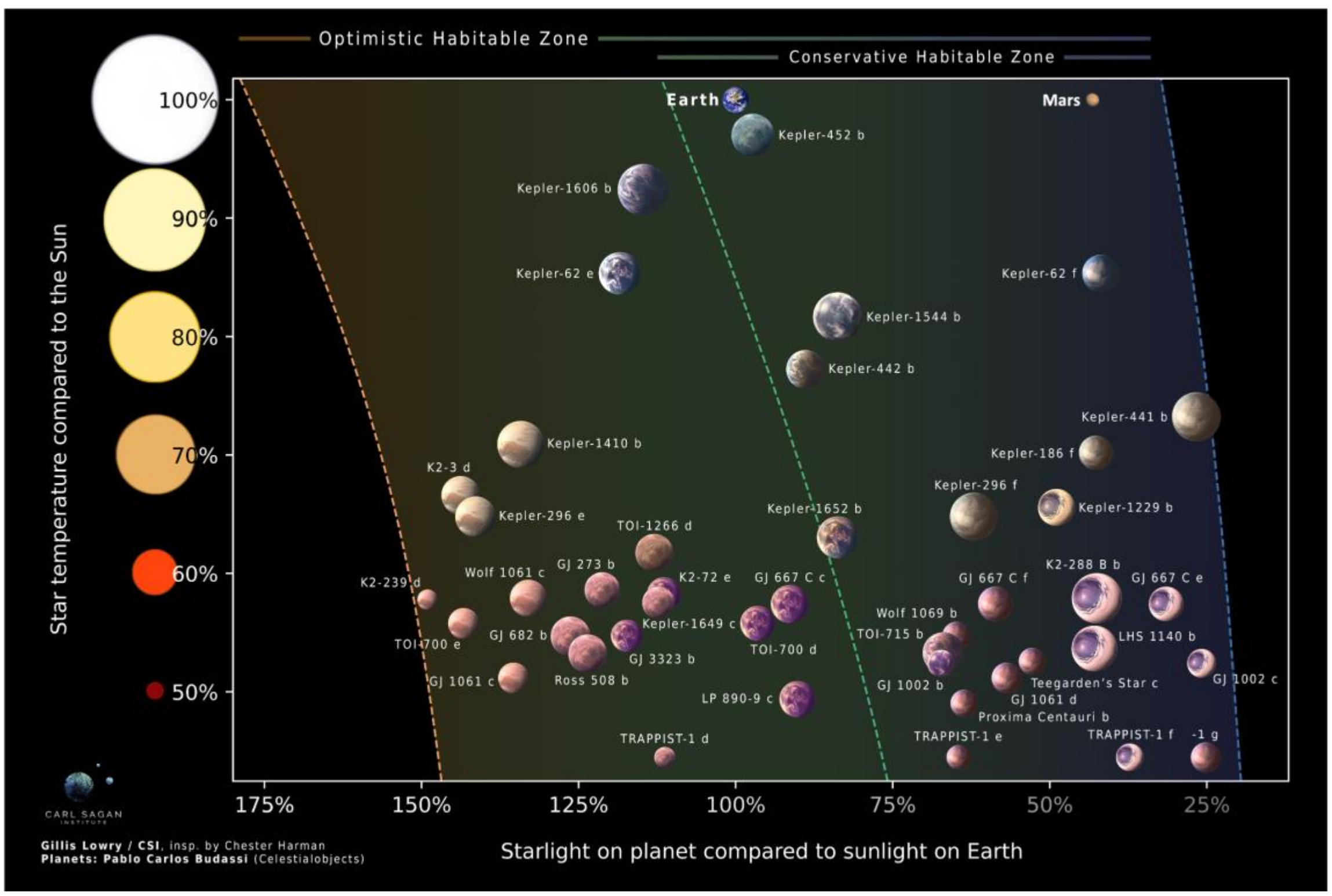

Figure 1: Potentially habitable planets presented in relation to a conservative and optimistic habitable zone boundary (Image Credit: G. Lowry). We have included Mars, which receives ~44% of the terrestrial insolation. Mars is scaled to match its relative size to Earth and is clearly within the "Conservative Habitable Zone" but has long since lost habitability.

While HWO will not achieve the fine spatial resolution of in situ spacecraft or orbiters, it will offer unparalleled spectral resolution and signal-to-noise, particularly in the ultraviolet, providing an essential complement to ongoing and future Mars missions. HWO's ability to observe the full Martian disk and exosphere in temporally resolved campaigns could be transformative for multi-scale atmospheric modeling and (via dayside aurora measurements) magnetospheric modeling, especially when used in conjunction with orbital datasets.

Beyond its scientific value, HWO’s observations could serve a dual role in supporting operational planning for robotic and human exploration. One speculative but compelling opportunity is the development of space weather and climate services for Mars. Global dust storms remain largely unpredictable beyond a few days, yet they pose significant risks to solar power generation, surface visibility, and potentially to landing systems for future crewed missions. HWO’s long-term monitoring of dust opacity, cloud formation, and atmospheric motion, particularly wind patterns, could help establish predictive thresholds for when regional dust events are likely to grow into planet encircling dust storms. Improved wind vector measurements from HWO observations would provide critical constraints on atmospheric models, enhancing their utility for short-term forecasting. In this way, HWO could lay the foundation for global Martian weather prediction, a direct benefit to mission planning and surface asset safety.

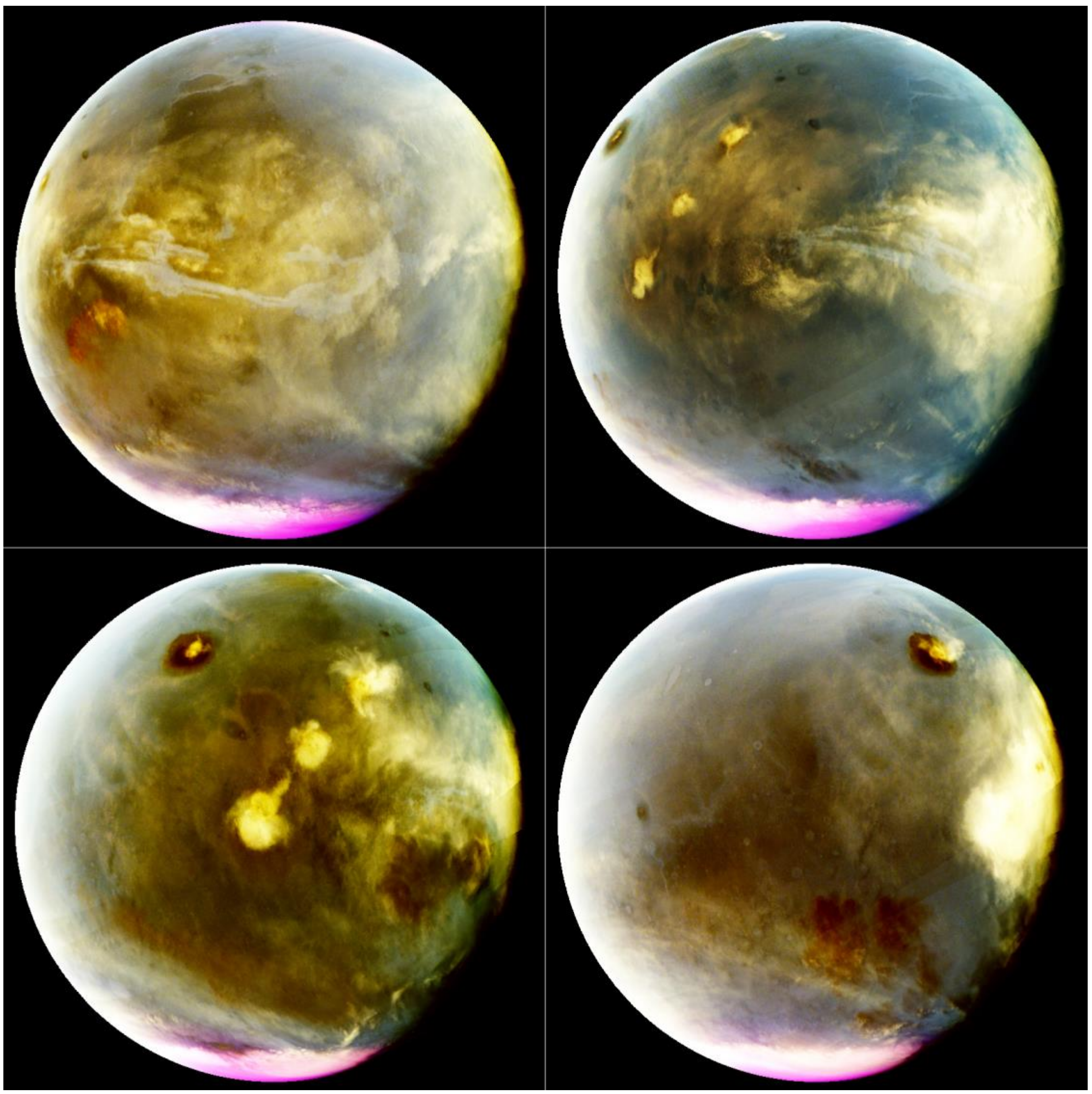

Figure 2: Mid-UV (200 - 340 nm) observations of Mars from the orbiting MAVEN spacecraft highlighting the presence of ozone (pink), clouds forming above the Tharsis volcano region (white) and regional dust events (reddish brown). While HWO may not achieve spatial resolutions at this scale, the increased spectral resolution and signal to noise will provide an invaluable resource to advance Mars science. (Image Credit: NASA, MAVEN, University of Colorado)

2. **Science Goal:** Quantify Mars as an analog to young habitable exoplanets and characterize the conditions for and processes leading to loss of habitability. Understanding present day Mars is a necessary prerequisite for building confidence in models that describe ancient Mars, a planet warm and wet enough to support rivers and lakes, even if only transiently, and by analogy, understand rocky exoplanets near the edge of the star's habitable zone, which might experience transient surface habitability, or subsurface habitability (i.e., in an ocean beneath a global ice layer as hypothesized for Kepler 62f and other "cold ocean" exoplanets).

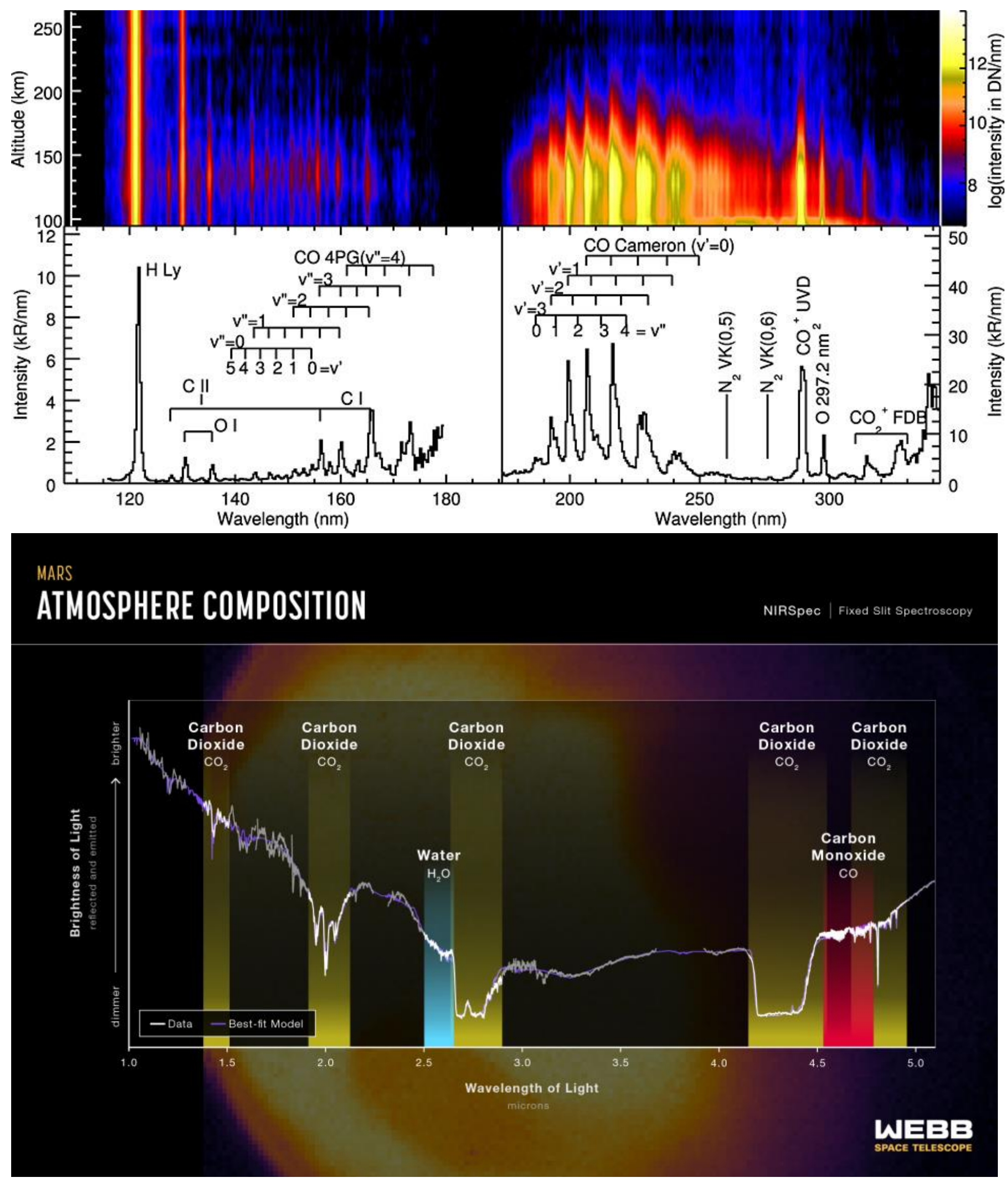

Figure 3: (Top) An observed spectrum of Mars in the UV from MAVEN, which demonstrates the variation in the airglow layer as a function of altitude, and the source of the emissions. (Image Credit: Jain et al. 2015) (Bottom) An observed spectrum of Mars compared to the best fit model in the IR from JWST, with several important molecular absorption features. (Image Credit: NASA, ESA, CSA, STScI, Mars JWST/GTO Team)

3. **Science Objectives:**

- Measure spatial and temporal variability in the upper atmosphere atomic D/H ratio and/or lower atmosphere H2O/HDO which serve as traces (and tracers) of atmospheric escape and water loss.
- Track clouds, ozone, and dust across Martian seasons and years to constrain drivers of the current Martian climate.
- Observe the Martian atmosphere from the UV through NIR to build comprehensive models of composition, circulation, and variability.

- Conduct short term campaigns (hours to days) for cloud and atmospheric motion tracking, and long-term campaigns (over months to years) to capture seasonal and interannual variations.
- Investigate coronal emissions of H, C and O atoms in the UV to quantify atmospheric loss to space.
- Constrain magnetospheric structure and processes via observations of dayside spectral signatures of precipitating particles, including solar-wind driven proton aurora and patchy proton aurora observable as Doppler-shifted H spectrum emission.

4. **Physical Parameters:**
    - Dayside atmospheric composition, including water vapor, ozone, isotopic ratios, as a function of latitude, longitude, local time, and season.
    - Temporal changes in clouds, dust, and ozone in reflected and emitted light across UV to IR wavelengths.
    - Full disk brightness and morphology of the disk, atmosphere and corona as Mars rotates (diurnal / longitudinal changes) and otherwise evolves with time (responding to solar input, dust activity, seasonal, and interannual variations).
    - Coronal hydrogen, carbon, oxygen emissions in the FUV and MUV.

5. **Description of Observations:** A UV/VIS Integral Field Spectrograph (IFS) on HWO offers an ideal platform for addressing these goals. A later generation NIR IFS would provide highly complementary data. Observations would include:

    - Full disk and coronal imaging spectroscopy of Mars during various phases of the Martian year. IFS mosaic-ing may enhance spatial resolution at the cost of temporal observation cadence of the full disk.
    - UV observations of thermospheric and coronal composition, and lower atmosphere ozone, ideal wavelength range: ~80 - 700 nm, minimum wavelength range: ~100 - 400 nm
    - NIR (up to 1.5 micron) observations for water vapor and isotopic ratios
    - Multiwavelength snapshots of Mars' dayside
    - Coordinated rapid response campaigns for events like dust storms or solar events
    - Contemporaneous or simultaneous multiband observations to characterize atmospheric motion and compositional changes.

6. **Impact on Requirements:** Designing HWO for Mars observations would have a modest impact on design requirements, with the main challenge being the brightness of the NUV + Visible dayside disk. As a superior planet, Mars observations are in general not as technically challenging as Venus, because they can be made at large solar elongation angles.

    - Visual Magnitude: -2.94 (closest) to +1.86 (farthest); Mean ~+0.71
    - Apparent Diameter: 3.5 to 25.1 arcsec (Mean ~15.5)

- FoR Requirements: Must accommodate full-disk and corona (~3+ $R_{mars}$ total)
- Solar Avoidance Constraints: Like Venus; must avoid observing too close to conjunction
- Spectral resolution requirements:
  - *Proton aurora and D/H measurements:* ~100 km/s at 121.6 nm for proton aurora observations (R~1000), ~5 km/s at 121.6 nm for upper atmosphere D/H observations (R~20,000), ~1 km/s at 121.6 nm for spectral separation and direct detection of escaping H component.
  - *Thermosphere and disk UV observations:* R~250 across 100-400 nm, consistent with MAVEN/IUVS level of detail
  - *Ozone:* R~250–500 across 200–300 nm to resolve the Hartley and Huggins absorption bands; higher resolution (~1000) desirable near 250 nm to separate ozone from dust opacity and to track seasonal column variations.
  - *H2O/HDO:* R~20,000–50,000 near 1.4 μm (or equivalent sensitivity in the 3.7 μm band) to spectrally resolve the HDO and H2O absorption features and measure isotopologue ratios; minimum R~5,000 for detection of the combined water vapor column.

7. **Relevance to NASA Strategic Goals** Mars exploration remains a key strategic goal for NASA over the next few decades, with a variety of planned and proposed SMD missions, as well as NASA's Moon to Mars architecture, which aims to increase effort to crewed space exploration, using the Moon as a short term goal in planning for crewed exploration of Mars. Leveraging HWO as a key part of this support network after several important observatories' mission lifetime has expired (Hubble, MAVEN, etc.) would be highly desirable for future human and robotic exploration.

   Mars science also directly aligns with key themes in the Origins, Worlds, and Life: The Planetary Science and Astrobiology Decadal Survey 2023-2032 (see Appendix).

**Impact on HWO Mission Requirements**

| ***Capability or Requirement*** | ***Necessary*** | ***Desired*** | ***Justification*** | ***Comments*** |
|---|---|---|---|---|
| UV observations | Yes, (100-400 nm) | Yes, (80-700 nm) | Observing exospheric H, C, O; Ozone, dust, and cloud tracking | Enables escape rate estimation, aurora detection and magnetosphere characterization, and climate modeling |
| Long wavelength observations (> 1.5 microns) | Yes | Yes | Detecting water vapor, HDO, and related | NIR complements UV tracing of the hydrological cycle |

| | | | molecular emissions | |
|---|---|---|---|---|
| Timing of observations | Yes | Freq. and duration will vary by campaign | Tracking cloud motion, auroral, and atmospheric changes | Requires both snapshot and long-baseline revisit cadence |
| High spatial resolution | Yes, can be achieved with mosaics | Ideally more pixels across the disk | Essential for resolving spatial variability from L2 | >1000 pixels across Mars' disk, ideally better if we want to resolve the corona and wind tracking |
| High spectral resolution | Yes | | Necessary to resolve molecular and isotopic lines | Should match or exceed exoplanet spectral resolution |
| Large field of view | Moderate | Yes | Capture full disk and exosphere during opposition | Disk + Corona span ~ 50 arcsec |
| Rapid response | Yes | | Reacting to transient phenomena (dust storms, solar flares, etc) | Requires coordination with solar monitoring observatories and triggering |

Appendix A.

Table A1. Themes consistent with investigations relevant to Mars exploration in Origins, Worlds, and Life: The Planetary Science and Astrobiology Decadal Survey 2023-2032 (OWL). Bold questions are particularly relevant in observing Mars.

| Q # | OWL Theme with Mars Relevance |
|---|---|
| 1.1 | What Were the Initial Conditions in the Solar System? |
| 1.2 | How Did Distinct Reservoirs of Gas and Solids Form and Evolve in the Protoplanetary Disk? |
| 1.3 | What Processes Led to the Production of Planetary Building Blocks? |
| **3.1** | How and When Did Asteroids and Inner Solar System Protoplanets Form? |
| **3.4** | What Processes Yielded Mars, Venus, and Mercury and Their Varied Initial States? |
| **3.6** | What Established the Primordial Inventories of Volatile Elements and Compounds in the Inner Solar System? |
| **4.2** | How Did Impact Bombardment Vary with Time and Location in the Solar System? |
| **4.4** | How Do the Physics and Mechanics of Impacts Produce Disruption of and Cratering on Planetary Bodies? |
| **5.1** | How Diverse Are the Compositions and Internal Structures Within and Among Solid Bodies? |
| **5.2** | How Have the Interiors of Solid Bodies Evolved? |
| **5.3** | How Have Surface/Near-Surface Characteristics and Compositions of Solid Bodies Been Modified by, and Recorded, Interior Processes? |
| **5.4** | How Have Surface Characteristics and Compositions of Solid Bodies Been Modified by, and Recorded, Surface Processes and Atmospheric Interactions? |
| **5.6** | What Drives Active Processes Occurring in the Interiors and on the Surfaces of Solid Bodies? |
| **6.1** | **How Do Solid-Body Atmospheres Form and What Was Their State During and Shortly after Accretion?** |
| **6.2** | **What Processes Govern the Evolution of Planetary Atmospheres and Climates Over Geologic Timescales?** |
| **6.3** | **What Processes Drive the Dynamics and Energetics of Atmospheres on Solid Bodies?** |
| **6.4** | **How Do Planetary Surfaces and Interiors Influence and Interact with Their Host Atmospheres?** |
| **6.5** | **What Processes Govern Atmospheric Loss to Space?** |
| **6.6** | **What Chemical and Microphysical Processes Govern the Clouds, Hazes, Chemistry and Trace Gas Composition of Solid Body Atmospheres?** |
| **10.1** | **What Is “Habitability”?** |
| 10.3 | **Water Availability: What Controls the Amount of Available Water on a Body Over Time?** |
| 10.5 | **What Is the Availability of Nutrients and Other Inorganic Ingredients to Support Life?** |
| 11.3 | **Life Detection: Is or Was There Life Elsewhere in the Solar System?** |
| 11.4 | Life Characterization: What Is the Nature of Life Elsewhere, If It Exists? |
| 12.1 | Evolution of the Protoplanetary Disk |
| 12.3 | **Origin of Earth and Inner Solar System Bodies** |
| 12.5 | Solid Body Interior and Surfaces |
| 12.6 | **Atmosphere and Climate Evolution on Solid Bodies** |
| 12.10 | **Dynamic Habitability** |
| 12.11 | **Search for Life Elsewhere** |

References:

Jain, S.K., Stewart, A.I.F., Schneider, N.M., Deighan, J., Stiepen, A., Evans, J.S., Stevens, M.H., Chaffin, M.S., Crismani, M., Mcclintock, W.E. and Clarke, J.T., 2015. The structure and variability of Mars upper atmosphere as seen in MAVEN/IUVS dayglow observations. Geophysical Research Letters, 42(21), pp.9023-9030.